\documentclass[preprint,5p,twocolumn,10pt]{elsarticle}

\usepackage[pdftex,hypertexnames=false,linktocpage=true]{hyperref}
\hypersetup{colorlinks=true,linkcolor=black,anchorcolor=darkgreen,citecolor=green!50!blue,filecolor=darkgreen,urlcolor=green,bookmarksnumbered=true,pdfview=FitB}

\usepackage{lipsum}
\usepackage{mathtools,cuted}

\usepackage{amsmath}
\usepackage{graphicx}
\usepackage{subfigure}
\usepackage{subcaption}

\usepackage{array}
\usepackage{booktabs}
\usepackage{arydshln}
\usepackage{multirow}

\newcommand{\eg}{\textit{e.g.}}

\usepackage{amsmath,amssymb,mathtools} 
\ifpdftex
    \usepackage{bbm}

\else
    \usepackage[bold-style=ISO]{unicode-math}
\fi

\usepackage{siunitx}  
\makeatletter
\newcommand{\vast}[1]{\bBigg@{#1}}
\makeatother

\newcommand{\intd}[3][]{\ifthenelse{\isempty{#3}}{\mathrm{d}^{#1} #2}{\frac{\mathrm{d}^{#1} #2}{#3}}\;}

\usepackage{stackengine}

\newcommand{\cond}{\; | \;}
\newcommand{\setcond}[2]{\ifthenelse{\isempty{#2}}{\{#1\}}{\{#1\cond{}#2\}}}

\usepackage{braket}
\usepackage{cancel}
\usepackage{braket}

\graphicspath{{../../plots/}}

\begin{document}

\begin{frontmatter}
\title{Pion to photon transition form factor: Beyond valence quarks}

\author[imp,ucas]{Xiaoyi Wu}
\ead{wuxiaoyi@impcas.ac.cn}

\author[imp,ucas]{Zhimin Zhu}
\ead{zhuzhimin@impcas.ac.cn}

\author[imp,ucas]{Ziyang Lin}
\ead{linzy@impcas.ac.cn}

\author[imp,ucas]{Chandan Mondal\corref{cor1}}
\ead{mondal@impcas.ac.cn}

\author[imp,ucas,affsch]{Jiangshan Lan\corref{cor1}}
\ead{jiangshanlan@impcas.ac.cn}

\author[imp,ucas]{Xingbo Zhao}
\ead{xbzhao@impcas.ac.cn}

\author[iowa]{James P. Vary}
\ead{jvary@iastate.edu}

\author[]{\\\vspace{0.2cm}(BLFQ Collaboration)}

\address[imp]{Institute of Modern Physics, Chinese Academy of Sciences, Lanzhou, Gansu, 730000, China}
\address[ucas]{School of Nuclear Physics, University of Chinese Academy of Sciences, Beijing, 100049, China}
\address[affsch]{Affiliated School of Huizhou University, Huizhou, Guangdong, 516001, China}
\address[iowa]{Department of Physics and Astronomy, Iowa State University, Ames, IA 50011, USA}
\cortext[cor1]{Corresponding author}

\begin{abstract}
We investigate the singly virtual transition form factor (TFF) for the $\pi^0\to\gamma^*\gamma$ process in the space-like region using the hard-scattering formalism within the Basis Light-Front Quantization (BLFQ) framework. This form factor is expressed in terms of the perturbatively calculable hard-scattering amplitudes (HSAs) and the light-front wave functions (LFWFs) of the pion. We obtain the pion LFWFs by diagonalizing the light-front QCD Hamiltonian, which is determined for its constituent quark-antiquark and quark-antiquark-gluon Fock sectors with a three-dimensional confinement. We employ the HSAs up to next-to-leading order (NLO) in the quark-antiquark Fock sector and leading order (LO) in the quark-antiquark-gluon Fock sector. The NLO correction to the TFF in the quark-antiquark Fock sector is of the same order as the LO contribution to the TFF in the quark-antiquark-gluon Fock sector. We find that while the quark-antiquark-gluon Fock sector has minimal effect in the large momentum transfer ($Q^2$) region, it has a noteworthy impact in the low-$Q^2$ region. Our results show that, after accounting for both Fock sectors, the TFF within the BLFQ framework aligns well with existing experimental data, particularly in the low $Q^2$ region.
\end{abstract}
\end{frontmatter}

\section{Introduction}\label{Sec1}
The structure of mesons remains a substantial puzzle in the current field of particle and nuclear physics~\cite{Accardi2012,Bacchetta2006,Belitsky2002,Lai2010,Pumplin2002}. 
Due to the difficulty in solving for their non-perturbative structure directly from QCD, it is challenging to gain theoretical insights into the internal structure of mesons. However, the development of the factorization theorem provides a pathway for studying this subject~\cite{Collins:2011zzd,Collins:1996fb,Diehl:2011yj,Collins:1985ue}. 
One of the simplest and most significant examples for factorization application in exclusive processes is the pion-photon transition form factor (TFF) with one real and one virtual photon, ${F}_{\pi\gamma}(Q^2)$. The TFF plays a crucial role in determining several important observables, for instance, the hadronic light-by-light contribution to the muon anomalous magnetic moment~\cite{Jegerlehner:2009ry,Nyffeler:2016gnb,Hoferichter:2018dmo}, and the rates of rare pseudoscalar decays~\cite{Hoferichter:2021lct,Husek:2015wta}.

Based on the factorization theorem, the TFF can be decomposed into the perturbatively calculable hard-scattering amplitudes (HSAs) and the light-front wave functions (LFWFs) which encode the non-perturbative dynamical information of the QCD bound states~\cite{Brodsky1980}. 
In 1980, Lepage and Brodsky first reported the leading-order (LO) HSAs for the TFF~\cite{Brodsky1980}. 
Subsequently, the next-to-leading order (NLO) correction has been examined in several studies~\cite{Braaten1982,delAguila1981,Kadantseva1985,Melic2001,Melic2002}. 
Meanwhile, the next-to-next-to order (NNLO) HSAs have been evaluated within the conformal scheme~\cite{Melic:2002ij,Braun:2021grd}, hard-collinear factorization theorem~\cite{Gao:2021iqq}, and principle of maximum commonality~\cite{Zhou:2023ivj}. 
On the other hand, despite the challenges posed by non-perturbative QCD dynamics, various phenomenological models and theoretical frameworks have been devised to compute the  distribution amplitudes.
These include methodologies like the chiral perturbation theory~\cite{Chang:2013pq}, QCD sum rules~\cite{Mikhailov:2016klg,Cheng:2020vwr,Stefanis:2020rnd}, Dyson-Schwinger and Bethe-Salpeter framework~\cite{Raya:2015gva,Raya:2016yuj}, AdS/QCD~\cite{Brodsky:2007hb}, and lattice QCD~\cite{RQCD:2019osh,Gao:2022vyh}.

Experimentally, the TFF has been extensively measured in space-like regions up to large transfer momentum, with several collaborations contributing measurements~\cite{CELLO_data,CLEO_data,BaBar_data,Belle_data}. 
In 2009, the BaBar Collaboration issued their data~\cite{BaBar_data} covering the kinematic range $Q^2\in[4,40] \; {\rm{GeV}}^2$, sparking a heated discussion due to its observation of non-conforming behavior at large $Q^2$, which contradicted the well-known asymptotic prediction~\cite{Brodsky1980}.
However, in 2012, the data from Belle Collaboration~\cite{Belle_data} showed that $Q^2F_{\pi\gamma}(Q^2)\rightarrow$ constant for $Q^2$ \textgreater 15 ${\rm{GeV}}^2$, which is consistent with the prediction of perturbative QCD~\cite{Brodsky1980}. 
The discrepancy between these two measurements in the large $Q^2$ region has consistently drawn the attention of both theoretical and experimental communities.

Our investigation of the TFF relies on the Basis Light-front Quantization (BLFQ) framework, offering a nonperturbative method for solving the relativistic many-body bound states problem within light-front quantum field theory~\cite{Vary:2009gt}.
Recent advancements in the BLFQ framework have led to successful applications in exploring various observables within QCD systems, \eg, the electromagnetic form factors and associated charge radii~\cite{Lan:2021wok,Mondal:2019jdg}, parton distribution functions~\cite{Lan:2021wok,Lan:2019img,Lan:2019rba,Lan:2019vui,Kaur:2024iwn,Mondal:2019jdg,Xu:2021wwj,Peng:2022lte}, transverse-momentum-dependent parton distributions~\cite{Zhu:2023lst,Kaur:2024iwn,Zhu:2023nhl,Hu:2022pgg,Hu:2022ctr} and generalized parton distributions~\cite{Adhikari:2021jrh,Zhang:2023xfe,Lin:2023ezw,Liu:2022fvl,Xu:2021wwj}. 
In an earlier exploration within the BLFQ framework, we reported the TFF only concerning the meson's valence quarks, where the Hamiltonian includes a three-dimensional confinement potential and the color-singlet Nambu--Jona-Lasinio interaction~\cite{Jia:2018ary,Chandon2021}. 
Our previous results show that valence Fock states alone are insufficient to describe TFF in a finite $Q^2$ region, suggesting that higher Fock states may have a significant impact.

In this work, we investigate the pion-photon TFF with one real and one virtual photon considering the valence Fock sector with constituent quark-antiquark and an
additional Fock sector, encompassing a quark-antiquark pair and a dynamical gluon of the pion.
Based on the BLFQ framework~\cite{Vary:2009gt}, we adopt the light-front QCD Hamiltonian that includes a model confining potential, and solve it in the first two Fock sectors, $|q\bar{q}\rangle$ and $|q\bar{q}g\rangle$~\cite{Lan:2021wok}. 
We evaluate the TFF through the overlap form of the HSAs and the LFWFs of the pion. 
The HSAs involving valence and valence-gluon Fock sectors are derived by the light-cone perturbation theory~\cite{Brodsky1980,Brodsky:1997de}. 
The LFWFs are obtained by diagonalizing the light-front Hamiltonian.

%=======================================================
\section{Meson LFWFs in the BLFQ framework\label{Sec2}}
%=======================================================
Following the Lepage-Brodsky convention~\cite{Brodsky1980}, the light-front variables are defined as $V^{\pm} \equiv V^0 \pm V^3$, and $\vec{V}_{\perp} \equiv (V_1,V_2)$. 
The wave functions of bound states are the eigenvectors of the light-front Hamiltonian equation: $P^+P^-|\Psi\rangle=M^2|\Psi\rangle$, where $P^+$ and $P^-$ represent the longitudinal momentum and the light-front Hamiltonian of the system, respectively. 
At the fixed light-front time, $x^+\equiv x^0+x^3$, the meson state with mass squared eigenvalue $M^2$ can be schematically expanded in terms of Fock sectors,
\begin{equation}
\begin{aligned}  |\Psi\rangle=\psi^{q\bar{q}}|q\bar{q}\rangle+\psi^{q\bar{q}g}|q\bar{q}g\rangle+\psi^{q\bar{q}q\bar{q}}|q\bar{q}q\bar{q}\rangle+\cdots,\label{psi0}
\end{aligned}
\end{equation}
where $\psi^{\cdots}$ are the LFWFs corresponding to the Fock sectors $|\cdots\rangle$. 
In this work, we truncate the infinite Fock expansion and retain the first two Fock sectors, which indicates that the meson is described by the quark-antiquark $\psi^{q\bar{q}}$ and quark-antiquark-gluon $\psi^{q\bar{q}g}$ LFWFs at the model scale. 

We adopt an effective light-front Hamiltonian, $P^-=P^-_{\rm{QCD}}+P^-_{\rm{C}}$~\cite{Lan:2021wok}, where $P^-_{\rm{QCD}}$ is the light-front QCD Hamiltonian, and $P^-_{\rm{C}}$ represents a model confining potential. 
In the light-front gauge $A^+=0$, the QCD Hamiltonian with one dynamical gluon is given by~\cite{Lan:2021wok,Brodsky:1997de},
\begin{equation}
\begin{aligned}
    P_{\rm{QCD}}^-&= \int \mathrm{d}x^- \mathrm{d}^2 x^{\perp} \Big\{\frac{1}{2}\bar{\psi}\gamma^+\frac{m_{0}^2+(i\partial^\perp)^2}{i\partial^+}\psi\\
    &+\frac{1}{2}A_a^i\left[m_g^2+(i\partial^\perp)^2\right] A^i_a +g_s\bar{\psi}\gamma_{\mu}T^aA_a^{\mu}\psi\\
    &+ \frac{1}{2}g_s^2\bar{\psi}\gamma^+T^a\psi\frac{1}{(i\partial^+)^2}\bar{\psi}\gamma^+T^a\psi \Big\},
\end{aligned}
\end{equation}
where $A^{\mu}_a$ and $\psi$ represent the gluon and the quark fields, respectively.
$T^a=\lambda^a/2$ is half of the Gell-Mann matrix. 
$g_s$ is the coupling constant. 
$m_g$ and $m_0$ stand for the model gluon mass and the bare quark mass, respectively. 
Though the gluon mass is zero in QCD, we artificially assign it as a phenomenologically motivated mass to fit the mass spectra in the low-energy scale~\cite{Lan:2021wok}. 
We introduce a mass counter term, $\delta m_q=m_0-m_q$, for the quark in the leading Fock sector to treat quark mass corrections from the higher Fock sectors, where $m_q$ is the renormalized quark mass. 
Following Ref.~\cite{Glazek:1992aq}, we allow an independent quark mass $m_f$ in the vertex interaction.

The confining potential consists of the transverse and longitudinal terms. 
In the valence Fock sector, it is written as~\cite{Lan:2021wok,Li:2015zda},
\begin{equation}
  \begin{aligned}\label{eqn:Hc}
  &P_{\rm{C}}^-P^+=\kappa^4\left\{x(1-x) \vec{r}_{\perp}^2-\frac{\partial_{x}[x(1-x)\partial_{x}]}{(m_q+m_{\bar{q}})^2}\right\},
  \end{aligned}
\end{equation}
where $\kappa$ is the strength coefficient of confinement, and $\vec r_{\perp}=\sqrt{x(1-x)}(\vec r_{q \perp}-\vec r_{\bar{q} \perp})$ represents the holographic variable~\cite{Brodsky:2014yha}. 
In the valence-gluon Fock sector, the confinement relies solely on the truncation of the BLFQ basis functions (see below). 

Using the BLFQ framework~\cite{Vary:2009gt}, we solve the Hamiltonian equation in a chosen basis space. 
The Fock sectors in Eq.~(\ref{psi0}) are taken to be direct products of single-particle states $|\alpha\rangle=\otimes_i|\alpha_i\rangle$. 
We employ the discretized light-cone quantization (DLCQ) basis~\cite{Brodsky:1997de} and a two-dimensional harmonic oscillator (2-D HO) basis functions to describe the longitudinal and transverse dynamics of single-particle states, respectively.
More specifically, in the longitudinal direction, the single particle is confined in a one-dimensional box of length $2L$ with periodic (anti-periodic) boundary conditions for the boson (fermion). 
For the $i^{\rm{th}}$ single-particle state, the longitudinal momentum is discretized as $p_i^+=\frac{2\pi}{L}k_i$, where the longitudinal quantum number $k_i$ is a half-integer (integer) for fermions (bosons).
Here, we neglect the zero mode $k_{i}=0$ for the boson. 
In the transverse plane, the 2-D HO wave function $\Phi_{n_im_i}(\vec{p}_{i\perp}, b)$ carries the radial and the angular quantum numbers denoted by $n_i$ and $m_i$, respectively. $\vec{p}_{i\perp}$ is the transverse momentum of the $i^{\rm{th}}$ particle. $b$ is the HO basis scale parameter, which defines the HO energy $E_{n_i,m_i}=(2n_i+|m_i|+1)b^2$.
Each single-particle state $|\alpha_i\rangle$ is characterized by four quantum numbers, $|\alpha_i\rangle=|k_i,n_i,m_i,\lambda_i\rangle$, where $\lambda_i$ is the light-front helicity.
In this work, the first two Fock sectors of the meson both allow for the unique color-singlet state. Therefore, there is no requirement for the color label to identify the color property of the LFWFs.

The projection of total angular momentum $M_J$ consisting of orbital angular momentum projection plus spin angular momentum projection is conserved in light-front field theory. 
Meanwhile, we respectively introduce truncations $N_{\rm{max}}$ and $K$ in transverse and longitudinal directions to perform the numerical calculation. 
These quantities satisfy the following conditions, 
\begin{equation}
\begin{aligned}
    \begin{cases}
          M_J=\sum_i (m_i+\lambda_i),\\
          N_{\rm{max}}\ge \sum_i(2n_i+|m_i|+1),\\
          K=\sum_{i}k_{i},
    \end{cases}
\end{aligned}
\end{equation}
where the dimensionless variable $K$ is used to parameterize the total longitudinal momentum $P^+$. 
For the $i^{\rm{th}}$ particle, the longitudinal momentum fraction is defined as $x_i=p_i^+/P^+=k_i/K$. 
The $N_{\rm{max}}$ truncation serves as the ultraviolet (UV) and infrared (IR) cutoffs. 
In momentum space, the UV cutoff $\Lambda_{\rm{UV}} \simeq b\sqrt{N_{\rm{max}}}$ and the IR cutoff $\Lambda_{\rm{IR}} \simeq b/\sqrt{N_{\rm{max}}}$ ~\cite{Zhao:2014xaa}.

Through diagonalizing the Hamiltonian, we directly obtain the mass spectra $M^2$ and the corresponding eigenvectors $\psi_{M_J}^{\mathcal{N}}(\{\alpha_i\})$, which can be converted to the LFWFs in momentum representation,
\begin{equation}
\begin{aligned}
  &\psi_{M_J,\{\lambda_i\}}^{\mathcal{N}}({\{x_i,\vec{p}_{i\perp}\}})\\
  &=\sum_{ \{n_i m_i\} }\psi_{M_J}^{\mathcal{N}}(\{\alpha_i\})\prod_{i=1}^{\mathcal{N}}  \Phi_{n_i m_i}(\vec{p}_{i\perp},b)\,,
\label{eqn:wf}
\end{aligned}
\end{equation}
where $\mathcal{N}$ is the total number of particles in the Fock sector. $\alpha_i$ represents the four quantum numbers $k_i$, $n_i$, $m_i$ and $\lambda_i$, characterizing the $i^{\rm{th}}$ particle as mentioned above.

With the truncation $\{N_{\text{max}}, K\}=\{14,15\}$, we determine the parameters summarized in Table~\ref{para} by fitting the mass eigenvalues of unflavored light mesons~\cite{Lan:2021wok}.
The LFWFs obtained by our method have simultaneously described the PDFs, decay constant, charge radii, and electromagnetic form factors for the light mesons, as well as the pion-nucleus induced $J/\psi$ production cross sections~\cite{Lan:2021wok}. 

%==================================================
\begin{table}[ht]\color{black}
  \caption{The model parameters obtained by fitting the mass eigenvalues of unflavored light mesons with the truncation $\{N_{\text{max}},K\}=\{14,15\}$~\cite{Lan:2021wok}.}
  \vspace{0.15cm}
  \label{para}
  \centering
    \setlength{\tabcolsep}{1mm}{
  \begin{tabular}{cccccc}
    \hline\hline
         $m_q$[GeV] & $m_g$[GeV] &$b$[GeV] &$\kappa$[GeV]  &${m}_f$[GeV]  &$g_s$ \\        
    \hline 
        0.39 & 0.60 &0.29 &0.65&5.69&1.92 \\       
    \hline\hline
  \end{tabular}}
\end{table}
%=====================================================

%========================================================
\section{Pion-photon transition form factors\label{Sec3}}
%========================================================
In general, there are two types of pion-photon transition form factors, specifically, the single and double virtual transition form factors extracted from the $\pi^0\rightarrow\gamma\gamma^*$ and $\pi^0\rightarrow\gamma^*\gamma^*$ processes, respectively~\cite{Chandon2021}. 
In this work, we limit our research to the single virtual pion-photon transition form factor $F_{\pi\gamma}(Q^2)$. The TFF is defined by the $\pi\gamma\gamma^\ast$ vertex~\cite{Brodsky1980},
\begin{equation}
\begin{aligned}\label{defination_TFF}
\Gamma_{\mu}=-ie^{2}F_{\pi\gamma}(Q^2)\epsilon_{\mu\nu\rho\sigma}p^{\nu}_{\pi}\varepsilon^{\ast\rho}q^{\sigma},
\end{aligned}
\end{equation}
where $q$ and $p_{\pi}$ are the momenta of the virtual photon and the pion, respectively. 
$\varepsilon$ is the polarization vector of the real photon. 
Following the prescription by Lepage and Brodsky~\cite{Brodsky1980}, we consider,
\begin{equation}
\begin{aligned}
\begin{cases}
\varepsilon^\ast&=(0,0,\vec{\varepsilon}_\perp^\ast),\vec{\varepsilon}_\perp^{\ast} \cdot \vec{q}_\perp=0,\\
p_\pi &\equiv (p^+,p^-,\vec{p}_\perp)=(1,m_\pi^2,\vec{0}),\\
q\ &=(0,\vec{q}_\perp^{2}-m_\pi^{2},\vec{q}_\perp) \Rightarrow q^2=-\vec{q}_\perp^2\equiv-Q^2.
\end{cases}
\end{aligned}
\end{equation}
In principle, the TFF includes the contribution from all Fock sectors. 
However, the contributions from higher Fock sectors are suppressed by powers of $1/Q$ compared to the valence Fock sector $|q\bar{q}\rangle$~\cite{Brodsky1980}, implying that higher Fock states may have a noticeable impact in regions where $Q$ is small.

Based on the factorization theorem, the TFF can be factorized into the HSAs and the LFWFs~\cite{Brodsky1980,Brodsky2011}.
Here, we evaluate the TFF from the convolution of the HSAs with the LFWFs considering our two chosen Fock sectors of the pion. The HSAs are derived by light-cone perturbation theory~\cite{Brodsky1980,Brodsky:1997de}, and the LFWFs are obtained by diagonalizing the light-front Hamiltonian~\cite{Lan:2021wok}.

 %==================================
\begin{figure}[ht]
  \centering
      \includegraphics[width=0.46\textwidth]{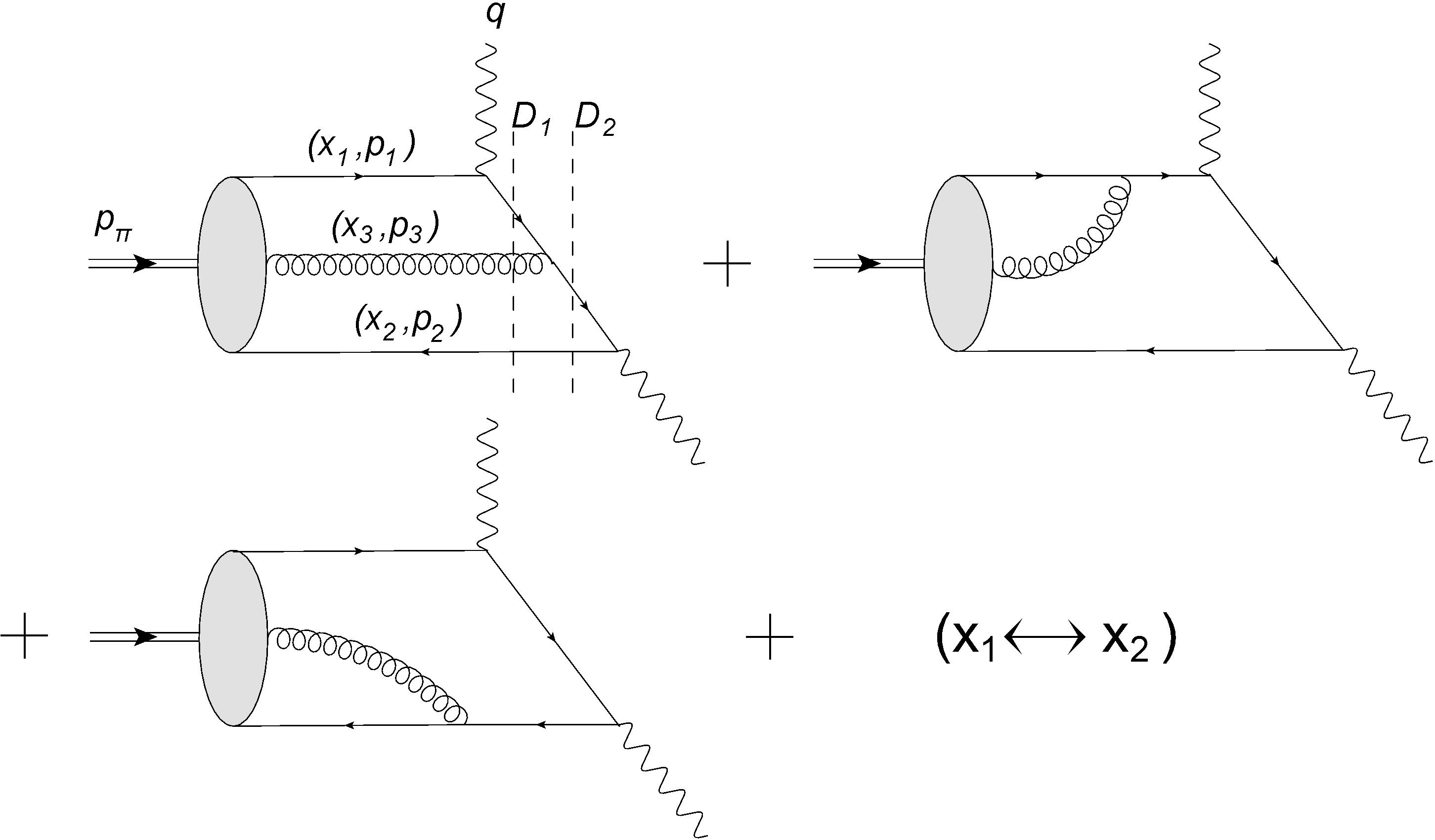}  
      \caption{The leading-order Feynman diagrams in the valence-gluon Fock sector related to $F_{\pi\gamma}$. $(x_1 \leftrightarrow x_2)$ represents the Feynman diagrams after exchanging the quark and antiquark.}
      \label{Feynman_diagram}
\end{figure}
%====================================

%===========================================
\subsection{TFF in the valence Fock sector}
%============================================
The TFF in the valence Fock sector is expressed as~\cite{Brodsky2011},
\begin{equation}\label{eq:TFF_convo}
\begin{aligned}
Q^2{F}^{q\bar{q}}_{\pi \gamma}(Q^2)&=\frac{4}{\sqrt{3}} \int_0^1 {\rm d}x \int^{\bar{x}Q}_0 \frac{{\rm d}^{2}\vec{p}_{\perp}}{16\pi^3}T^{q\bar{q}}(x,Q^2) \psi_{\lambda_1\lambda_2}^2(x,\vec{p}_{\perp}),
\end{aligned}
\end{equation}
where $\psi^2_{\lambda_1\lambda_2}$ is the LFWF of the valence Fock sector with
$\lambda_1$ and $\lambda_2$ being the light-front helicities of the quark and antiquark, respectively. 
Since the pion is spinless, we omit the subscript $M_J$. 
The HSA up to NLO in the valence Fock sector, $T^{q\bar{q}}(x,Q^2)$, is given by~\cite{Braaten1982,delAguila1981,Kadantseva1985,Melic2001,Melic2002},
\begin{equation}
\begin{aligned}\label{eq:THNLO}
&T^{q\bar{q}}(x,Q^2) =\frac{1}{1-x}\\
&+\frac{\alpha_s(\mu_{\mathrm R})}{4 \pi}
C_{\mathrm F} \frac{1}{1-x} \,\Big[-9-\frac{1-x}{x} {\rm{ln}}(1-x) \\
& + {\rm{ln}}^2(1-x)   +\left\{ 3+2 \,{\rm{ln}}(1-x) \right\} {\rm{ln}} \left(\frac{Q^2}{\mu^2_{\mathrm R}} \right)
\Big],
\end{aligned}
\end{equation}
where $\alpha_s(\rm \mu_R)=\frac{4\pi}{\beta_{0}\rm{ln}(\mu^{2}_{R}/\Lambda^2_{\rm{QCD}})}$ is the strong running coupling constant with $\beta_{0}=\frac{11}{3}C_{\rm A}-\frac{2}{3}n_{\rm f}$ and $\Lambda_{\rm{QCD}}=225$ MeV. 
Here, the color factors are given by $C_{\rm F}=\frac{4}{3}$ and $C_{\rm A}=3$. 
The number of active flavors is $n_{\rm f}=3$. 
For simplicity, we adopt the regularization scale as $\mu^2_{\rm R}=Q^2$ to eliminate the large logarithm term. 
The first term in Eq.~(\ref{eq:THNLO}) is the LO HSA. 
The term related to $\alpha_s$ is the NLO correction of the HSA in the valence Fock sector.

%=================================================
\subsection{TFF in the $|q\bar{q}g\rangle$ Fock sector}
%=================================================
The NLO correction to the TFF from the valence Fock sector is of the same order as the LO contribution from the valence-gluon Fock sector. 
The LO contribution from the valence-gluon Fock sector to the TFF can be expressed as 
\begin{align}
    Q^2{F}^{q\bar{q}g}_{\pi \gamma}(Q^2)&=\frac{8m_q\sqrt{2\pi\alpha_{s}}}{3Q^2} \sum_{i=1}^3 \int_{0}^1 {\rm d}x_1 \int_0^{1-x_1} {\rm d}x_2 \nonumber\\ 
    &\times \int^{f_i(x_1,x_2,Q)}_0 \frac{{\rm d}^{2}\vec{p}_{1\perp}}{16\pi^3} \int^{g_i(x_1,x_2,Q)}_0 \frac{{\rm d}^{2}\vec{p}_{2\perp}}{16\pi^3} \nonumber\\  
    &\times T^{q\bar{q}g}_i(x_1,x_2)\psi_{\lambda_1\lambda_2\lambda_3}^3(x_1,\vec{p}_{1\perp},x_2,\vec{p}_{2\perp})\nonumber\\ 
    &+(x_1 \longleftrightarrow x_2),\label{Eq:TFF_qqg}
\end{align}
where  $\psi_{\lambda_1\lambda_2\lambda_3}^3$ is the LFWF of the valence-gluon Fock sector with $\lambda_1$, $\lambda_2$, and $\lambda_3$ being the light-front helicities of the quark, antiquark, and gluon, respectively. 
$T_i^{q\bar{q}g}$ denotes the HSA corresponding to the $i^{\rm{th}}$ diagram among the first three diagrams in Fig.~\ref{Feynman_diagram}.
Due to the charge conjugation invariance, the LFWFs of the pion and the HSAs exhibit symmetry under the exchange of the quark and antiquark. 
Consequently, evaluating the first term in Eq.~(\ref{Eq:TFF_qqg}) suffices, implying that the total TFF in the valence-gluon Fock sector is twice the magnitude of the first term.
The explicit expressions of LO HSAs in the valence-gluon Fock sector read as,
\begin{align}
  T^{q\bar{q}g}_1(x_1,x_2)&= \frac{x_1}{(1-x_1)(1-x_2)x^{2}_{2}}\nonumber\\
  &+\frac{1-x_1-x_2}{(1-x_1)(1-x_2)x_{2}},\label{HSAs.qqg1}\\
  T^{q\bar{q}g}_2(x_1,x_2)&=\frac{x_1+x_2-1}{x_1x_2(1-x_2)},\label{HSAs.qqg2}\\
  T^{q\bar{q}g}_3(x_1,x_2)&=\frac{2(x_1+x_2-1)}{x_1(1-x_2)^2}.\label{HSAs.qqg3}
\end{align}
The upper limits of the integral over transverse momenta in Eq.~\eqref{Eq:TFF_qqg} are given by
\begin{align}
  f_1&=\rm{min}(\frac{1-x_1}{2},\frac{1-x_1-x_2}{1-x_2})Q,\label{lim1}\\
  g_1&=\rm{min}(x_2,\frac{1-x_1-x_2}{x_1},\sqrt{\frac{x_2(1-x_1-x_2)}{x_1}})Q,\label{lim2}\\
  f_2&=Q\; , \; g_2=\rm{min}(x_2,\sqrt{x_2(1-x_2)})Q,\label{lim3}\\
  f_3&=\rm{min}(1-x_1,\sqrt{x_1(1-x_1)})Q\; , \; g_3=Q.\label{lim4}
\end{align}
The detailed derivation for the TFF with LO HSAs in $|q\bar{q}g\rangle$ Fock sector is provided in the~\ref{appendix_A}.
%==================================================
\begin{figure*}[ht]
  \centering
      \includegraphics[width=3.15in,height=2.6in]{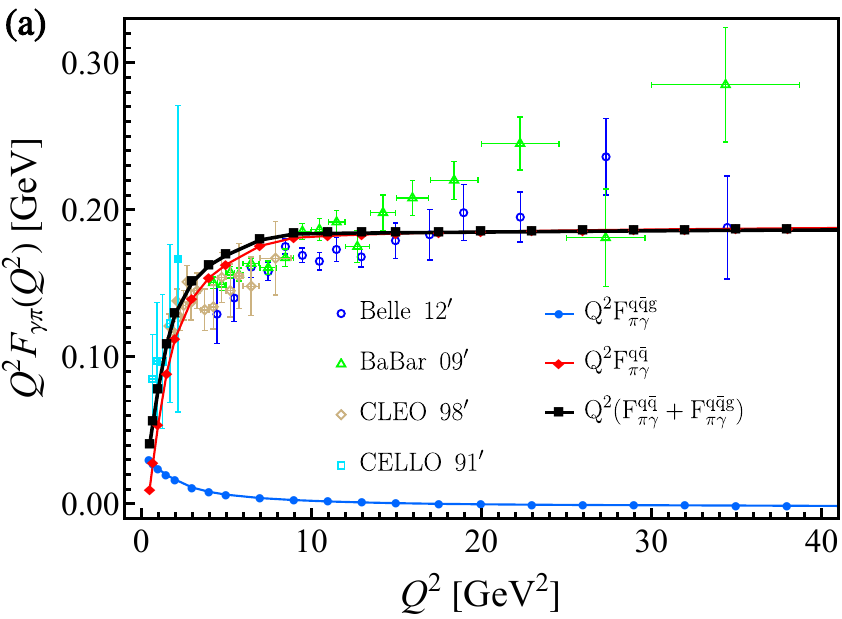}
      \qquad 
      \includegraphics[width=3.15in,height=2.6in]{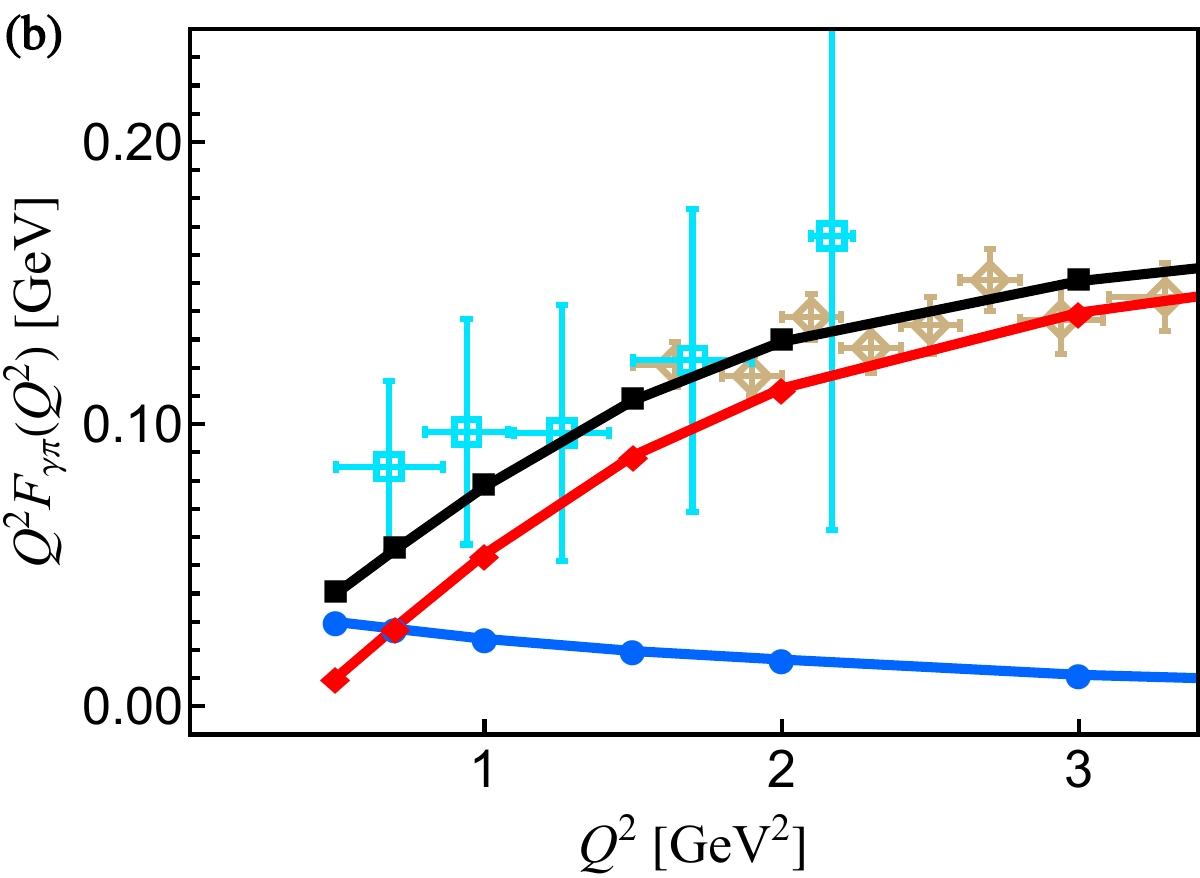}
      \caption{The BLFQ results for the $\pi^{0}\rightarrow\gamma^{\ast}\gamma$ transition form factor with truncation parameters $N_{\text{max}} = 14$ and $K = 15$: (a) the TFF in full $Q^2$ region; (b) the TFF in low-$Q^2$ region. The blue and red curves correspond to the results with the LO HSA in the $|q\bar{q}g\rangle$ Fock sector and the NLO HSA in the $|q\bar{q}\rangle$ Fock sector, respectively. The black curve represents the sum of them. Data are taken from CELLO~\cite{CELLO_data}, CLEO~\cite{CLEO_data}, Belle~\cite{Belle_data}, and BaBar~\cite{BaBar_data} Collaborations.}
      % \caption{\textcolor{black}{(a) shows the BLFQ results for the $\pi^{0}\rightarrow\gamma^{\ast}\gamma$ transition form factor with truncation parameters $N_{\text{max}}=14$ and $K=15$. The blue and red curves correspond to the results with the LO HSA in the $|q\bar{q}g\rangle$ Fock sector and the NLO HSA in the $|q\bar{q}\rangle$ Fock sector, respectively. The black curve represents the sum of them. 
      % Data are taken from CELLO~\cite{CELLO_data}, CLEO~\cite{CLEO_data}, Belle~\cite{Belle_data}, and BABAR~\cite{BaBar_data} Collaborations. (b) highlights the TFF at low-$Q^2$ region.}}
      \label{Fig_1_of_tff}
\end{figure*}
%==================================================
%=======================================
\section{Numerical results\label{Sec4}}
%=======================================
For our numerical calculations, we set the truncation limits to $N_{\rm max} =14$ for the transverse direction and  $K=15$ for the longitudinal direction. With the model parameters summarized in Table~\ref{para}, we obtain the LFWFs of the pion in the first two Fock sectors with its mass $M=0.139 \; \rm{GeV}$ by solving the light-front Hamiltonian equation. At the model scale, the probabilities of the pion being in the Fock states $|q\bar{q}\rangle$ and $|q\bar{q}g\rangle$ are $50.8\%$ and $49.2\%$ respectively. Subsequently, we compute the pion to photon TFF by convoluting the HSAs with the pion LFWFs, as described in Eqs.~\eqref{eq:TFF_convo} and \eqref{Eq:TFF_qqg}.

Figure~\ref{Fig_1_of_tff} shows our results for the pion to photon TFF, $Q^{2}F_{\pi\gamma}(Q^{2})$, within the BLFQ framework and compares  with the available experimental data from the CELLO~\cite{CELLO_data}, CLEO~\cite{CLEO_data}, Belle~\cite{Belle_data} and BaBar Collaborations~\cite{BaBar_data}. 
The red and blue curves represent the contributions from the $|q\bar{q}\rangle$ and $|q\bar{q}g\rangle$ Fock sectors, respectively, with the black curve corresponding to the total TFF. The $\chi^2$ per degree of freedom is 4.1 over all data included in Fig.~2a. Excluding the BaBar data lowers this to 2.4 while additionally excluding the Belle data, leaving only the CLEO and CELLO data in Fig. 2b, lowers it to 1.1. We find a good consistency between our results and the data reported by CELLO~\cite{CELLO_data}, CLEO~\cite{CLEO_data}, Belle Collaborations~\cite{Belle_data}. However, our result deviates from the rapid growth in the large $Q^2$ region reported by the BaBar Collaboration~\cite{BaBar_data}. Some theoretical studies suggest that the BaBar data are incompatible with QCD calculations~\cite{Zhou:2023ivj,Ahmady:2022dfv,Chandon2021,Cao:2021ddi,Gao:2021iqq,Stefanis:2020rnd,Mikhailov:2009sa,Roberts:2010rn,Bakulev:2011rp,Wu:2011gf}. Conversely, there are phenomenological studies that reproduce the BaBar data for the pion to photon TFF~\cite{Polyakov:2009je,Brodsky:2011yv,Wu:2010zc,Kroll:2010bf,RuizArriola:2010mrb,Gorchtein:2011vf,Pham:2011zi,Dorokhov:2010zzb,Agaev:2010aq,Kotko:2009ij}.

As $Q^2$ increases, the magnitude of $Q^2F^{q\bar{q}}_{\pi\gamma}$ gradually increases and eventually saturates, while that of $Q^2F^{q\bar{q}g}_{\pi\gamma}$ gradually decreases and eventually vanishes. This behavior is consistent with the scaling predicted by pQCD~\cite{Brodsky1980, Braaten1982}. Specifically, the TFF follows that $Q^2F_{\pi\gamma} (Q^2\to\infty) = 2f_\pi$, where $f_\pi$ is the decay constant~\cite{Brodsky2011}. Additionally, contributions to the TFF from higher Fock sectors are suppressed by powers of $1/Q$. Note that the second Fock sector includes a dynamical gluon, introducing an additional propagator at the corresponding HSA. This additional propagator generates a factor of $Q^2$ in the denominator. Consequently, $F^{q\bar{q}g}_{\pi\gamma}$ exhibits a lower power of $Q^2$ compared to $F^{q\bar{q}}_{\pi\gamma}$, as indicated by the comparison between Eqs.~\eqref{eq:TFF_convo} and \eqref{Eq:TFF_qqg}. The contribution from the $|q\bar{q}g\rangle$ Fock sector decreases more rapidly and has negligible influence in the large $Q^2$ region. However, the contribution from the $|q\bar{q}g\rangle$ Fock component has a visible impact in the small $Q^{2}$ region, which improves the theoretical description of experimental data for $Q^{2}<5$ ${\rm{GeV}}^{2}$ as seen in Fig. \textcolor{blue}{2(b)}.

\begin{figure}[ht]
  \centering
      \includegraphics[width=3.15in,height=2.6in]{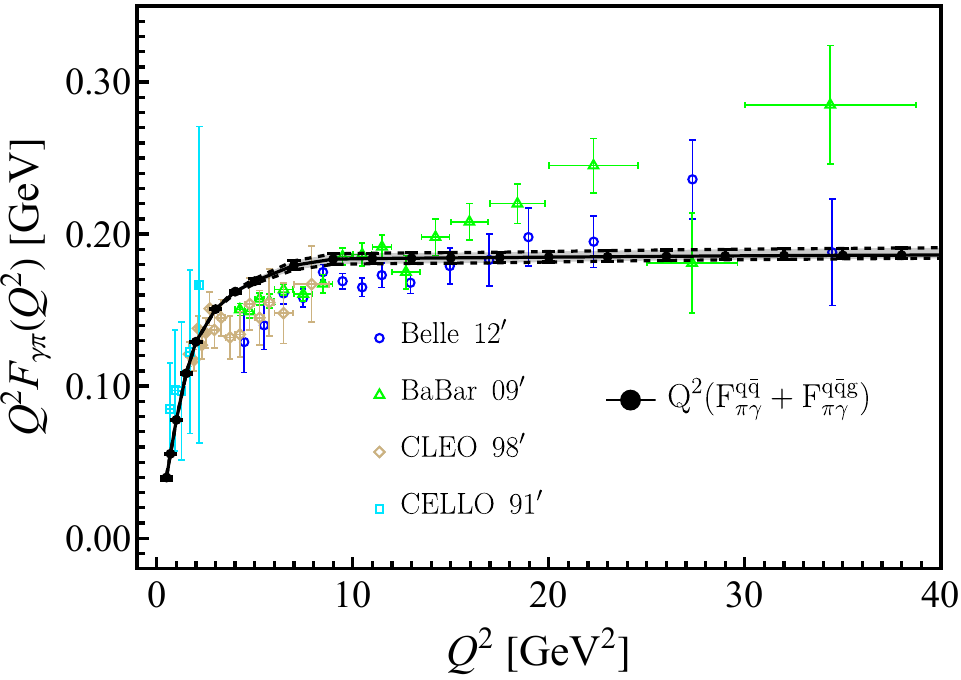}  
      \caption{Uncertainty estimation for the TFF, $Q^2(F^{q\bar{q}}_{\pi\gamma}+F^{q\bar{q}g}_{\pi\gamma})$. The uncertainties are quantified by comparing the results at different model quark masses.}
      \label{fig_uncertainty}
\end{figure}

In this work, we quantify the uncertainties by comparing the results at different quark masses without changing the other parameters. More specifically, we shift the model quark mass from 0.39 GeV to 0.37 GeV and to 0.41 GeV, and then solve the light-front stationary Schrödinger equation to obtain the two Fock-state LFWFs of the pion. Finally, we evaluate the TFF at different quark masses to quantify the uncertainties. We present the result, including the uncertainties, for $Q^2(F^{q\bar{q}}_{\pi\gamma}+F^{q\bar{q}g}_{\pi\gamma})$ in Fig.~\ref{fig_uncertainty}. The largest relative uncertainty is 4$\%$ at $Q^2 = 35 \ \rm{GeV}^2$, which demonstrates the stability of the model results with respect to slight changes in the input parameters.
%==============================
\section{Summary\label{Sec6}}
%==============================
We evaluated the pion to photon transition form factor (TFF) with one on-shell and one off-shell photon using the hard-scattering formalism within the Basis Light-Front Quantization (BLFQ) framework. The TFF was obtained by convoluting the light-front wave functions of the pion in its first two Fock sectors, $|q\bar{q}\rangle$ and $|q\bar{q}g\rangle$, with the hard-scattering amplitudes (HSAs) derived using light-cone perturbation theory. The pion's wave functions were determined from the eigenvectors of the light-front QCD Hamiltonian relevant to constituent quark-antiquark and quark-antiquark-gluon
Fock components of the mesons in the light-cone gauge  along with a three-dimensional confinement in the leading Fock sector. Notably, our results are purely predictive, with no parameter adjustments made when calculating the TFF. We found that the contribution from the valence Fock sector dominates in the large-$Q^2$ region, consistent with the scaling behavior predicted by perturbative QCD. The contribution from the $|q\bar{q}g\rangle$ Fock sector has a modest impact in the low-$Q^2$ region, enhancing the description of the experimental data compared to considering only the leading Fock sector. Overall, our results are in good agreement with the experimental data, particularly from the Belle, CLEO, and CELLO collaborations. Combining this success in the TFF with achievements in light meson spectroscopy, decay constants, electromagnetic form factors, and parton distribution functions, the BLFQ framework provides a natural approach to describing the relativistic structure of the pion.

\section*{Acknowledgements}
We thank Yang Li, Siqi Xu, Jiatong Wu, and Jialin Chen for many helpful discussions. 
Z. Zhu is supported by the Natural Science Foundation of Gansu Province, China, Grant No. 23JRRA571.
C. M. is supported by new faculty start up funding by the Institute of Modern Physics, Chinese Academy of Sciences, Grant No. E129952YR0.
J. Lan is supported by the Special Research Assistant Funding Project, Chinese Academy of Sciences, by the National Science Foundation of Gansu Province, China, Grant No. 23JRRA631, and National Natural Science Foundation of China under Grant No. 12305095.
X. Z. is supported by new faculty startup funding by the Institute of Modern Physics, Chinese Academy of Sciences, by Key Research Program of Frontier Sciences, Chinese Academy of Sciences, Grant No. ZDBS-LY-7020, by the Foundation for Key Talents of Gansu Province, by the Central Funds Guiding the Local Science and Technology Development of Gansu Province, Grant No. 22ZY1QA006, by Gansu International Collaboration and Talents Recruitment Base of Particle Physics (2023-2027), by International Partnership Program of the Chinese Academy of Sciences, Grant No. 016GJHZ2022103FN, by National Natural Science Foundation of China, Grant No.~12375143, by National Key R\&D Program of China, Grant No.~2023YFA1606903 and by the Strategic Priority Research Program of the Chinese Academy of Sciences, Grant No.~XDB34000000.
J. P. V. is supported by the Department of Energy under Grant No. DE-SC0023692. 
A portion of the computational resources were also provided by Advanced Computing Center in Taiyuan and by Sugon Computing Center in Xi'an.

%==================================================================
\appendix
\section{TFF with LO HSAs in $|q\bar{q}g\rangle$ Fock sector}\label{appendix_A}
%==================================================================
Here, we provide a detailed derivation of the HSAs in the quark-antiquark-gluon Fock sector, as given in Eqs.~\eqref{HSAs.qqg1}-\eqref{HSAs.qqg3}, and explain the reasons for the upper limit of the integral specified in Eqs.~\eqref{lim1}-\eqref{lim4}.
In our BLFQ framework for the pion, the S-wave dominates the LFWFs of the $|q\bar{q}g\rangle$  Fock sector, accounting for 98.4\% of the probability in this sector. Consequently, we disregard the contributions of the P- and D-waves from this Fock sector.
For the S-wave in this Fock sector, the convolution of the LFWFs with the HSAs is invariant under helicity flip. Considering this symmetry, we only need to derive the contribution from one type of S-wave helicity state.

Taking the first diagram in Fig.~\ref{Feynman_diagram} as an example, we set the helicity of the quark, antiquark and gluon as $\lambda_1$, $\lambda_2$, and $\lambda_3$, respectively.
Based on the light-cone perturbation theory, the corresponding convolution form of the TFF in the $|q\bar{q}g\rangle$ Fock sector with $\{\lambda_1,\lambda_2,\lambda_3\} = \{\uparrow,\uparrow,\downarrow\}$ can be expressed as,
\begin{align}
    &F_{\pi\gamma,1\mathrm{st}}^{q\bar{q}g}=\frac{\sqrt{4\pi\alpha_s(n^2_{\rm{c}}-1)}(e^2_u-e^2_d)}{i\sqrt{2}(\vec{\varepsilon}^{~\ast}_\perp \times \vec{q}_\perp)} \int^1_0 [{\rm d}x] \int^{\infty}_0 [{\rm d}p_\perp] \frac{1}{D_1} \nonumber\\ 
    \times &\frac{1}{D_2}\sum_{\lambda\lambda^\prime} \Bigg{[} \frac{\bar{v}_{\lambda_2}(x_2,\vec{p}_{2\perp})\gamma^{\rho}\varepsilon^\ast_{\rho}u_{\lambda}(x_1+x_3,\vec{p}_{1\perp}+\vec{p}_{3\perp}+\vec{q}_\perp)}{\sqrt{x_2}\sqrt{x_1+x_3}} \nonumber\\ 
    \times &\frac{\bar{u}_{\lambda}(x_1+x_3,\vec{p}_{1\perp}+\vec{p}_{3\perp}+\vec{q}_{\perp})\gamma^\nu\varepsilon_\nu(\lambda_3)u_{\lambda^\prime}(x_1,\vec{p}_{1\perp}+\vec{q}_{\perp})}{\sqrt{x_1}\sqrt{x_1+x_3}} \nonumber\\
    \times &\frac{\bar{u}_{\lambda^\prime}(x_1,\vec{p}_{1\perp}+\vec{q}_{\perp})\gamma^+u_{\lambda_1}(x_1,\vec{p}_{1\perp})}{\sqrt{x_1}\sqrt{x_1}} \Bigg{]} \psi_{\lambda_1\lambda_2\lambda_3}^3,
\end{align}
where $[{\rm d}x]\equiv{\rm d}x_1\,{\rm d}x_2\,{\rm d}x_3\,{\rm \delta}(1-x_1-x_2-x_3)$ and $[{\rm d}p_\perp]\equiv\frac{{\rm d}^2\vec{p}_{1\perp}}{16\pi^3}\frac{{\rm d}^2\vec{p}_{2\perp}}{16\pi^3}$. 
$\varepsilon^\ast$ and $\varepsilon$ are the polarization vectors of the final (on-shell) photon and the gluon, respectively.
$e_{u}$ and $e_d$ are the quark charges in units of $e$, $\alpha_s$ is the strong running coupling constant, and $n_{\rm{c}}=3$ is the number of colors. 
$\psi_{\lambda_1\lambda_2\lambda_3}^3$ represents the LFWF corresponding to the helicity state we chose above. 
Here, we omit the arguments $(x_1,\vec{p}_{1\perp},x_2,\vec{p}_{2\perp})$ from the LFWF, where the longitudinal momentum fraction and the transverse momentum of the gluon are $x_3=1-x_1-x_2$ and $\vec{p}_{3\perp} = -\vec{p}_{1\perp}-\vec{p}_{2\perp}$, respectively.
$D_1$ and $D_2$ are the energy denominators.
For the first diagram, they are written as, 
\begin{align}
      D_1&= \vec{q}_\perp^2-\frac{(\vec{p}_{1\perp}+\vec{q}_\perp)^2+m_q^2}{x_1}-\frac{\vec{p}^2_{2\perp}+m_q^2}{x_2}-\frac{\vec{p}_{3\perp}^2}{x_3}, \\
      D_2&= \vec{q}_{\perp}^2-\frac{(\vec{p}_{1\perp}+\vec{p}_{3\perp}+\vec{q}_\perp)^2+m_q^2}{x_1+x_3}-\frac{\vec{p}^2_{2\perp}+m_q^2}{x_2}.
  \end{align}

Substituting in the expression of the spinors and the polarization vector of the gluon, and neglecting the quark mass $m_q$ relative to $Q^2$, the TFF corresponding to the first diagram becomes 
\begin{align}
    &F_{\pi\gamma,1\mathrm{st}}^{q\bar{q}g}=\frac{8m_q\sqrt{2\pi\alpha_{s}}}{3\vec{\varepsilon}^{~\ast}_\perp \times \vec{q}_\perp}\int^1_0 [\mathrm{d}x] \int^{\infty}_0 [\mathrm{d}p_\perp] \nonumber\\ 
    &\times \frac{1}{\vec{q}_{\perp}^2-\frac{(\vec{p}_{1\perp}+\vec{q}_\perp)^2}{x_1}-\frac{\vec{p}^2_{2\perp}}{x_2}-\frac{\vec{p}_{3\perp}^2}{x_3}}\nonumber\\%\frac{x_2(1-x_2)}{(x_2\vec{q}_\perp-\vec{p}_{2\perp})^2} \nonumber\\ 
    &\times \frac{1}{\vec{q}_\perp^2-\frac{(\vec{p}_{1\perp}+\vec{p}_{3\perp}+\vec{q}_\perp)^2}{x_1+x_3}-\frac{\vec{p}_{2\perp}^2}{x_2}}\nonumber\\
    &\times\Bigg{[}\frac{\vec{\varepsilon}_{\perp}^{~\ast} \times ((1-x_1-x_2)\vec{q}_{\perp}+(1-x_2)\vec{p}_{1\perp}+x_1\vec{p}_{2\perp})}{x_2(1-x_2)^2(x_1+x_2-1)} \nonumber\\
    &+\frac{\vec{\varepsilon}_{\perp}^{~\ast} \times (\vec{p}_{2\perp}-x_2\vec{q}_\perp)x_3}{x_1x_2(x_1+x_3)^2}\Bigg{]}\psi_{\lambda_1=\uparrow, \lambda_2=\uparrow,\lambda_3=\downarrow}^3.
    \label{eq:TFF_qqg1}
\end{align}

The S-wave component is concentrated in the low transverse momentum region because a bound state has minimal amplitude when its components move with large transverse momentum.
We follow the technique prescribed in Ref.~\cite{Brodsky1980} to simplify the energy denominators and the expressions in the square brackets in Eq.~(\ref{eq:TFF_qqg1}).
We take the small transverse momentum limit, setting $|\vec{p}_{1\perp}| \le f_1=\rm{min}(\frac{1-x_1}{2},\frac{1-x_1-x_2}{1-x_2})Q$ and $|\vec{p}_{2\perp}| \le g_1=\rm{min}(x_2,\frac{1-x_1-x_2}{x_1},\sqrt{\frac{x_2(1-x_1-x_2)}{x_1}})Q$.
Consequently, the energy denominators are simplified to $D_1=D_2=\vec{q}_\perp^2$, and only the terms related to $\vec{q}_\perp$ in the square brackets survive. Ultimately, by substituting $\vec{q}^2_\perp=Q^2$, the expression of the TFF is reduced to Eq.~\eqref{Eq:TFF_qqg} with the HSA $T_1^{q\bar{q}g}$ in Eq.~(\ref{HSAs.qqg1}).  The HSAs corresponding to the other diagrams given in Eqs.~(\ref{HSAs.qqg2}) and (\ref{HSAs.qqg3}), can similarly be derived using this approach.

\biboptions{sort&compress}
\bibliographystyle{elsarticle-num}
\bibliography{PionGamma_TFF_Ref.bib}

\begin{thebibliography}{10}
\expandafter\ifx\csname url\endcsname\relax
  \def\url#1{\texttt{#1}}\fi
\expandafter\ifx\csname urlprefix\endcsname\relax\def\urlprefix{URL }\fi
\expandafter\ifx\csname href\endcsname\relax
  \def\href#1#2{#2} \def\path#1{#1}\fi

\bibitem{Accardi2012}
A.~Accardi, et~al., {Electron Ion Collider: The Next QCD Frontier}: {Understanding the glue that binds us all}, Eur. Phys. J. A 52~(9) (2016) 268.
\newblock \href {http://arxiv.org/abs/1212.1701} {\path{arXiv:1212.1701}}, \href {https://doi.org/10.1140/epja/i2016-16268-9} {\path{doi:10.1140/epja/i2016-16268-9}}.

\bibitem{Bacchetta2006}
A.~Bacchetta, M.~Diehl, K.~Goeke, A.~Metz, P.~J. Mulders, M.~Schlegel, {Semi-inclusive deep inelastic scattering at small transverse momentum}, JHEP 02 (2007) 093.
\newblock \href {http://arxiv.org/abs/hep-ph/0611265} {\path{arXiv:hep-ph/0611265}}, \href {https://doi.org/10.1088/1126-6708/2007/02/093} {\path{doi:10.1088/1126-6708/2007/02/093}}.

\bibitem{Belitsky2002}
A.~V. Belitsky, X.~Ji, F.~Yuan, {Final state interactions and gauge invariant parton distributions}, Nucl. Phys. B 656 (2003) 165--198.
\newblock \href {http://arxiv.org/abs/hep-ph/0208038} {\path{arXiv:hep-ph/0208038}}, \href {https://doi.org/10.1016/S0550-3213(03)00121-4} {\path{doi:10.1016/S0550-3213(03)00121-4}}.

\bibitem{Lai2010}
H.-L. Lai, M.~Guzzi, J.~Huston, Z.~Li, P.~M. Nadolsky, J.~Pumplin, C.~P. Yuan, {New parton distributions for collider physics}, Phys. Rev. D 82 (2010) 074024.
\newblock \href {http://arxiv.org/abs/1007.2241} {\path{arXiv:1007.2241}}, \href {https://doi.org/10.1103/PhysRevD.82.074024} {\path{doi:10.1103/PhysRevD.82.074024}}.

\bibitem{Pumplin2002}
J.~Pumplin, D.~R. Stump, J.~Huston, H.~L. Lai, P.~M. Nadolsky, W.~K. Tung, {New generation of parton distributions with uncertainties from global QCD analysis}, JHEP 07 (2002) 012.
\newblock \href {http://arxiv.org/abs/hep-ph/0201195} {\path{arXiv:hep-ph/0201195}}, \href {https://doi.org/10.1088/1126-6708/2002/07/012} {\path{doi:10.1088/1126-6708/2002/07/012}}.

\bibitem{Collins:2011zzd}
J.~Collins, {Foundations of Perturbative QCD}, Vol.~32 of Cambridge Monographs on Particle Physics, Nuclear Physics and Cosmology, Cambridge University Press, 2023.
\newblock \href {https://doi.org/10.1017/9781009401845} {\path{doi:10.1017/9781009401845}}.

\bibitem{Collins:1996fb}
J.~C. Collins, L.~Frankfurt, M.~Strikman, {Factorization for hard exclusive electroproduction of mesons in QCD}, Phys. Rev. D 56 (1997) 2982--3006.
\newblock \href {http://arxiv.org/abs/hep-ph/9611433} {\path{arXiv:hep-ph/9611433}}, \href {https://doi.org/10.1103/PhysRevD.56.2982} {\path{doi:10.1103/PhysRevD.56.2982}}.

\bibitem{Diehl:2011yj}
M.~Diehl, D.~Ostermeier, A.~Schafer, {Elements of a theory for multiparton interactions in QCD}, JHEP 03 (2012) 089, [Erratum: JHEP 03, 001 (2016)].
\newblock \href {http://arxiv.org/abs/1111.0910} {\path{arXiv:1111.0910}}, \href {https://doi.org/10.1007/JHEP03(2012)089} {\path{doi:10.1007/JHEP03(2012)089}}.

\bibitem{Collins:1985ue}
J.~C. Collins, D.~E. Soper, G.~F. Sterman, {Factorization for Short Distance Hadron - Hadron Scattering}, Nucl. Phys. B 261 (1985) 104--142.
\newblock \href {https://doi.org/10.1016/0550-3213(85)90565-6} {\path{doi:10.1016/0550-3213(85)90565-6}}.

\bibitem{Jegerlehner:2009ry}
F.~Jegerlehner, A.~Nyffeler, {The Muon g-2}, Phys. Rept. 477 (2009) 1--110.
\newblock \href {http://arxiv.org/abs/0902.3360} {\path{arXiv:0902.3360}}, \href {https://doi.org/10.1016/j.physrep.2009.04.003} {\path{doi:10.1016/j.physrep.2009.04.003}}.

\bibitem{Nyffeler:2016gnb}
A.~Nyffeler, {Precision of a data-driven estimate of hadronic light-by-light scattering in the muon $g-2$: Pseudoscalar-pole contribution}, Phys. Rev. D 94~(5) (2016) 053006.
\newblock \href {http://arxiv.org/abs/1602.03398} {\path{arXiv:1602.03398}}, \href {https://doi.org/10.1103/PhysRevD.94.053006} {\path{doi:10.1103/PhysRevD.94.053006}}.

\bibitem{Hoferichter:2018dmo}
M.~Hoferichter, B.-L. Hoid, B.~Kubis, S.~Leupold, S.~P. Schneider, {Pion-pole contribution to hadronic light-by-light scattering in the anomalous magnetic moment of the muon}, Phys. Rev. Lett. 121~(11) (2018) 112002.
\newblock \href {http://arxiv.org/abs/1805.01471} {\path{arXiv:1805.01471}}, \href {https://doi.org/10.1103/PhysRevLett.121.112002} {\path{doi:10.1103/PhysRevLett.121.112002}}.

\bibitem{Hoferichter:2021lct}
M.~Hoferichter, B.-L. Hoid, B.~Kubis, J.~L\"udtke, {Improved Standard-Model prediction for $\pi^0\to e^+e^-$}, Phys. Rev. Lett. 128~(17) (2022) 172004.
\newblock \href {http://arxiv.org/abs/2105.04563} {\path{arXiv:2105.04563}}, \href {https://doi.org/10.1103/PhysRevLett.128.172004} {\path{doi:10.1103/PhysRevLett.128.172004}}.

\bibitem{Husek:2015wta}
T.~Husek, S.~Leupold, {Two-hadron saturation for the pseudoscalar\textendash{}vector\textendash{}vector correlator and phenomenological applications}, Eur. Phys. J. C 75~(12) (2015) 586.
\newblock \href {http://arxiv.org/abs/1507.00478} {\path{arXiv:1507.00478}}, \href {https://doi.org/10.1140/epjc/s10052-015-3778-x} {\path{doi:10.1140/epjc/s10052-015-3778-x}}.

\bibitem{Brodsky1980}
G.~P. Lepage, S.~J. Brodsky, Exclusive processes in perturbative quantum chromodynamics, Phys. Rev. D 22 (1980) 2157--2198.
\newblock \href {https://doi.org/10.1103/PhysRevD.22.2157} {\path{doi:10.1103/PhysRevD.22.2157}}.

\bibitem{Braaten1982}
E.~Braaten, {QCD CORRECTIONS TO MESON - PHOTON TRANSITION FORM-FACTORS}, Phys. Rev. D 28 (1983) 524.
\newblock \href {https://doi.org/10.1103/PhysRevD.28.524} {\path{doi:10.1103/PhysRevD.28.524}}.

\bibitem{delAguila1981}
F.~del Aguila, M.~K. Chase, {HIGHER ORDER QCD CORRECTIONS TO EXCLUSIVE TWO PHOTON PROCESSES}, Nucl. Phys. B 193 (1981) 517--528.
\newblock \href {https://doi.org/10.1016/0550-3213(81)90344-8} {\path{doi:10.1016/0550-3213(81)90344-8}}.

\bibitem{Kadantseva1985}
E.~P. Kadantseva, S.~V. Mikhailov, A.~V. Radyushkin, {Total $\alpha^- s$ Corrections to Processes $\gamma^* \gamma^* \to \pi^0$ and $\gamma^* \pi \to \pi$ in a Perturbative {QCD}}, Yad. Fiz. 44 (1986) 507--516.

\bibitem{Melic2001}
B.~Melic, B.~Nizic, K.~Passek, {BLM scale setting for the pion transition form-factor}, Phys. Rev. D 65 (2002) 053020.
\newblock \href {http://arxiv.org/abs/hep-ph/0107295} {\path{arXiv:hep-ph/0107295}}, \href {https://doi.org/10.1103/PhysRevD.65.053020} {\path{doi:10.1103/PhysRevD.65.053020}}.

\bibitem{Melic2002}
B.~Melic, D.~Mueller, K.~Passek-Kumericki, {Next-to-next-to-leading prediction for the photon to pion transition form-factor}, Phys. Rev. D 68 (2003) 014013.
\newblock \href {http://arxiv.org/abs/hep-ph/0212346} {\path{arXiv:hep-ph/0212346}}, \href {https://doi.org/10.1103/PhysRevD.68.014013} {\path{doi:10.1103/PhysRevD.68.014013}}.

\bibitem{Melic:2002ij}
B.~Melic, D.~Mueller, K.~Passek-Kumericki, {Next-to-next-to-leading prediction for the photon to pion transition form-factor}, Phys. Rev. D 68 (2003) 014013.
\newblock \href {http://arxiv.org/abs/hep-ph/0212346} {\path{arXiv:hep-ph/0212346}}, \href {https://doi.org/10.1103/PhysRevD.68.014013} {\path{doi:10.1103/PhysRevD.68.014013}}.

\bibitem{Braun:2021grd}
V.~M. Braun, A.~N. Manashov, S.~Moch, J.~Schoenleber, {Axial-vector contributions in two-photon reactions: Pion transition form factor and deeply-virtual Compton scattering at NNLO in QCD}, Phys. Rev. D 104~(9) (2021) 094007.
\newblock \href {http://arxiv.org/abs/2106.01437} {\path{arXiv:2106.01437}}, \href {https://doi.org/10.1103/PhysRevD.104.094007} {\path{doi:10.1103/PhysRevD.104.094007}}.

\bibitem{Gao:2021iqq}
J.~Gao, T.~Huber, Y.~Ji, Y.-M. Wang, {Next-to-Next-to-Leading-Order QCD Prediction for the Photon-Pion Form Factor}, Phys. Rev. Lett. 128~(6) (2022) 062003.
\newblock \href {http://arxiv.org/abs/2106.01390} {\path{arXiv:2106.01390}}, \href {https://doi.org/10.1103/PhysRevLett.128.062003} {\path{doi:10.1103/PhysRevLett.128.062003}}.

\bibitem{Zhou:2023ivj}
H.~Zhou, J.~Yan, Q.~Yu, X.-G. Wu, {Updated determination of the pion-photon transition form factor}, Phys. Rev. D 108~(7) (2023) 074020.
\newblock \href {http://arxiv.org/abs/2306.10510} {\path{arXiv:2306.10510}}, \href {https://doi.org/10.1103/PhysRevD.108.074020} {\path{doi:10.1103/PhysRevD.108.074020}}.

\bibitem{Chang:2013pq}
L.~Chang, I.~C. Cloet, J.~J. Cobos-Martinez, C.~D. Roberts, S.~M. Schmidt, P.~C. Tandy, {Imaging dynamical chiral symmetry breaking: pion wave function on the light front}, Phys. Rev. Lett. 110~(13) (2013) 132001.
\newblock \href {http://arxiv.org/abs/1301.0324} {\path{arXiv:1301.0324}}, \href {https://doi.org/10.1103/PhysRevLett.110.132001} {\path{doi:10.1103/PhysRevLett.110.132001}}.

\bibitem{Mikhailov:2016klg}
S.~V. Mikhailov, A.~V. Pimikov, N.~G. Stefanis, {Systematic estimation of theoretical uncertainties in the calculation of the pion-photon transition form factor using light-cone sum rules}, Phys. Rev. D 93~(11) (2016) 114018.
\newblock \href {http://arxiv.org/abs/1604.06391} {\path{arXiv:1604.06391}}, \href {https://doi.org/10.1103/PhysRevD.93.114018} {\path{doi:10.1103/PhysRevD.93.114018}}.

\bibitem{Cheng:2020vwr}
S.~Cheng, A.~Khodjamirian, A.~V. Rusov, {Pion light-cone distribution amplitude from the pion electromagnetic form factor}, Phys. Rev. D 102~(7) (2020) 074022.
\newblock \href {http://arxiv.org/abs/2007.05550} {\path{arXiv:2007.05550}}, \href {https://doi.org/10.1103/PhysRevD.102.074022} {\path{doi:10.1103/PhysRevD.102.074022}}.

\bibitem{Stefanis:2020rnd}
N.~G. Stefanis, {Pion-photon transition form factor in light cone sum rules and tests of asymptotics}, Phys. Rev. D 102~(3) (2020) 034022.
\newblock \href {http://arxiv.org/abs/2006.10576} {\path{arXiv:2006.10576}}, \href {https://doi.org/10.1103/PhysRevD.102.034022} {\path{doi:10.1103/PhysRevD.102.034022}}.

\bibitem{Raya:2015gva}
K.~Raya, L.~Chang, A.~Bashir, J.~J. Cobos-Martinez, L.~X. Guti\'errez-Guerrero, C.~D. Roberts, P.~C. Tandy, {Structure of the neutral pion and its electromagnetic transition form factor}, Phys. Rev. D 93~(7) (2016) 074017.
\newblock \href {http://arxiv.org/abs/1510.02799} {\path{arXiv:1510.02799}}, \href {https://doi.org/10.1103/PhysRevD.93.074017} {\path{doi:10.1103/PhysRevD.93.074017}}.

\bibitem{Raya:2016yuj}
K.~Raya, M.~Ding, A.~Bashir, L.~Chang, C.~D. Roberts, {Partonic structure of neutral pseudoscalars via two photon transition form factors}, Phys. Rev. D 95~(7) (2017) 074014.
\newblock \href {http://arxiv.org/abs/1610.06575} {\path{arXiv:1610.06575}}, \href {https://doi.org/10.1103/PhysRevD.95.074014} {\path{doi:10.1103/PhysRevD.95.074014}}.

\bibitem{Brodsky:2007hb}
S.~J. Brodsky, G.~F. de~Teramond, {Light-Front Dynamics and AdS/QCD Correspondence: The Pion Form Factor in the Space- and Time-Like Regions}, Phys. Rev. D 77 (2008) 056007.
\newblock \href {http://arxiv.org/abs/0707.3859} {\path{arXiv:0707.3859}}, \href {https://doi.org/10.1103/PhysRevD.77.056007} {\path{doi:10.1103/PhysRevD.77.056007}}.

\bibitem{RQCD:2019osh}
G.~S. Bali, V.~M. Braun, S.~B\"urger, M.~G\"ockeler, M.~Gruber, F.~Hutzler, P.~Korcyl, A.~Sch\"afer, A.~Sternbeck, P.~Wein, {Light-cone distribution amplitudes of pseudoscalar mesons from lattice QCD}, JHEP 08 (2019) 065, [Addendum: JHEP 11, 037 (2020)].
\newblock \href {http://arxiv.org/abs/1903.08038} {\path{arXiv:1903.08038}}, \href {https://doi.org/10.1007/JHEP08(2019)065} {\path{doi:10.1007/JHEP08(2019)065}}.

\bibitem{Gao:2022vyh}
X.~Gao, A.~D. Hanlon, N.~Karthik, S.~Mukherjee, P.~Petreczky, P.~Scior, S.~Syritsyn, Y.~Zhao, {Pion distribution amplitude at the physical point using the leading-twist expansion of the quasi-distribution-amplitude matrix element}, Phys. Rev. D 106~(7) (2022) 074505.
\newblock \href {http://arxiv.org/abs/2206.04084} {\path{arXiv:2206.04084}}, \href {https://doi.org/10.1103/PhysRevD.106.074505} {\path{doi:10.1103/PhysRevD.106.074505}}.

\bibitem{CELLO_data}
H.~J. Behrend, et~al., {A Measurement of the pi0, eta and eta-prime electromagnetic form-factors}, Z. Phys. C 49 (1991) 401--410.
\newblock \href {https://doi.org/10.1007/BF01549692} {\path{doi:10.1007/BF01549692}}.

\bibitem{CLEO_data}
J.~Gronberg, et~al., {Measurements of the meson - photon transition form-factors of light pseudoscalar mesons at large momentum transfer}, Phys. Rev. D 57 (1998) 33--54.
\newblock \href {http://arxiv.org/abs/hep-ex/9707031} {\path{arXiv:hep-ex/9707031}}, \href {https://doi.org/10.1103/PhysRevD.57.33} {\path{doi:10.1103/PhysRevD.57.33}}.

\bibitem{BaBar_data}
B.~Aubert, et~al., {Measurement of the gamma gamma* ---\ensuremath{>} pi0 transition form factor}, Phys. Rev. D 80 (2009) 052002.
\newblock \href {http://arxiv.org/abs/0905.4778} {\path{arXiv:0905.4778}}, \href {https://doi.org/10.1103/PhysRevD.80.052002} {\path{doi:10.1103/PhysRevD.80.052002}}.

\bibitem{Belle_data}
S.~Uehara, et~al., {Measurement of $\gamma \gamma^* \to \pi^0$ transition form factor at Belle}, Phys. Rev. D 86 (2012) 092007.
\newblock \href {http://arxiv.org/abs/1205.3249} {\path{arXiv:1205.3249}}, \href {https://doi.org/10.1103/PhysRevD.86.092007} {\path{doi:10.1103/PhysRevD.86.092007}}.

\bibitem{Vary:2009gt}
J.~P. Vary, H.~Honkanen, J.~Li, P.~Maris, S.~J. Brodsky, A.~Harindranath, G.~F. de~Teramond, P.~Sternberg, E.~G. Ng, C.~Yang, {Hamiltonian light-front field theory in a basis function approach}, Phys. Rev. C 81 (2010) 035205.
\newblock \href {http://arxiv.org/abs/0905.1411} {\path{arXiv:0905.1411}}, \href {https://doi.org/10.1103/PhysRevC.81.035205} {\path{doi:10.1103/PhysRevC.81.035205}}.

\bibitem{Lan:2021wok}
J.~Lan, K.~Fu, C.~Mondal, X.~Zhao, j.~P. Vary, {Light mesons with one dynamical gluon on the light front}, Phys. Lett. B 825 (2022) 136890.
\newblock \href {http://arxiv.org/abs/2106.04954} {\path{arXiv:2106.04954}}, \href {https://doi.org/10.1016/j.physletb.2022.136890} {\path{doi:10.1016/j.physletb.2022.136890}}.

\bibitem{Mondal:2019jdg}
C.~Mondal, S.~Xu, J.~Lan, X.~Zhao, Y.~Li, D.~Chakrabarti, J.~P. Vary, {Proton structure from a light-front Hamiltonian}, Phys. Rev. D 102~(1) (2020) 016008.
\newblock \href {http://arxiv.org/abs/1911.10913} {\path{arXiv:1911.10913}}, \href {https://doi.org/10.1103/PhysRevD.102.016008} {\path{doi:10.1103/PhysRevD.102.016008}}.

\bibitem{Lan:2019img}
J.~Lan, C.~Mondal, M.~Li, Y.~Li, S.~Tang, X.~Zhao, J.~P. Vary, {Parton Distribution Functions of Heavy Mesons on the Light Front}, Phys. Rev. D 102~(1) (2020) 014020.
\newblock \href {http://arxiv.org/abs/1911.11676} {\path{arXiv:1911.11676}}, \href {https://doi.org/10.1103/PhysRevD.102.014020} {\path{doi:10.1103/PhysRevD.102.014020}}.

\bibitem{Lan:2019rba}
J.~Lan, C.~Mondal, S.~Jia, X.~Zhao, J.~P. Vary, {Pion and kaon parton distribution functions from basis light front quantization and QCD evolution}, Phys. Rev. D 101~(3) (2020) 034024.
\newblock \href {http://arxiv.org/abs/1907.01509} {\path{arXiv:1907.01509}}, \href {https://doi.org/10.1103/PhysRevD.101.034024} {\path{doi:10.1103/PhysRevD.101.034024}}.

\bibitem{Lan:2019vui}
J.~Lan, C.~Mondal, S.~Jia, X.~Zhao, J.~P. Vary, {Parton Distribution Functions from a Light Front Hamiltonian and QCD Evolution for Light Mesons}, Phys. Rev. Lett. 122~(17) (2019) 172001.
\newblock \href {http://arxiv.org/abs/1901.11430} {\path{arXiv:1901.11430}}, \href {https://doi.org/10.1103/PhysRevLett.122.172001} {\path{doi:10.1103/PhysRevLett.122.172001}}.

\bibitem{Kaur:2024iwn}
S.~Kaur, J.~Wu, Z.~Hu, J.~Lan, C.~Mondal, X.~Zhao, J.~P. Vary, {Quark and gluon distributions in \ensuremath{\rho}-meson from basis light-front quantization}, Phys. Lett. B 851 (2024) 138563.
\newblock \href {http://arxiv.org/abs/2401.03480} {\path{arXiv:2401.03480}}, \href {https://doi.org/10.1016/j.physletb.2024.138563} {\path{doi:10.1016/j.physletb.2024.138563}}.

\bibitem{Xu:2021wwj}
S.~Xu, C.~Mondal, J.~Lan, X.~Zhao, Y.~Li, J.~P. Vary, {Nucleon structure from basis light-front quantization}, Phys. Rev. D 104~(9) (2021) 094036.
\newblock \href {http://arxiv.org/abs/2108.03909} {\path{arXiv:2108.03909}}, \href {https://doi.org/10.1103/PhysRevD.104.094036} {\path{doi:10.1103/PhysRevD.104.094036}}.

\bibitem{Peng:2022lte}
T.~Peng, Z.~Zhu, S.~Xu, X.~Liu, C.~Mondal, X.~Zhao, J.~P. Vary, {Basis light-front quantization approach to \ensuremath{\Lambda} and \ensuremath{\Lambda}c and their isospin triplet baryons}, Phys. Rev. D 106~(11) (2022) 114040.
\newblock \href {http://arxiv.org/abs/2208.00355} {\path{arXiv:2208.00355}}, \href {https://doi.org/10.1103/PhysRevD.106.114040} {\path{doi:10.1103/PhysRevD.106.114040}}.

\bibitem{Zhu:2023lst}
Z.~Zhu, Z.~Hu, J.~Lan, C.~Mondal, X.~Zhao, J.~P. Vary, {Transverse structure of the pion beyond leading twist with basis light-front quantization}, Phys. Lett. B 839 (2023) 137808.
\newblock \href {http://arxiv.org/abs/2301.12994} {\path{arXiv:2301.12994}}, \href {https://doi.org/10.1016/j.physletb.2023.137808} {\path{doi:10.1016/j.physletb.2023.137808}}.

\bibitem{Zhu:2023nhl}
Z.~Zhu, T.~Peng, Z.~Hu, S.~Xu, C.~Mondal, X.~Zhao, J.~P. Vary, {Transverse momentum structure of strange and charmed baryons: A light-front Hamiltonian approach}, Phys. Rev. D 108~(3) (2023) 036009.
\newblock \href {http://arxiv.org/abs/2304.05058} {\path{arXiv:2304.05058}}, \href {https://doi.org/10.1103/PhysRevD.108.036009} {\path{doi:10.1103/PhysRevD.108.036009}}.

\bibitem{Hu:2022pgg}
Z.~Hu, S.~Xu, C.~Mondal, X.~Zhao, J.~P. Vary, {BLFQ calculations of the proton leading twist quark TMDs}, Rev. Mex. Fis. Suppl. 3~(3) (2022) 0308101.
\newblock \href {https://doi.org/10.31349/SuplRevMexFis.3.0308101} {\path{doi:10.31349/SuplRevMexFis.3.0308101}}.

\bibitem{Hu:2022ctr}
Z.~Hu, S.~Xu, C.~Mondal, X.~Zhao, J.~P. Vary, {Transverse momentum structure of proton within the basis light-front quantization framework}, Phys. Lett. B 833 (2022) 137360.
\newblock \href {http://arxiv.org/abs/2205.04714} {\path{arXiv:2205.04714}}, \href {https://doi.org/10.1016/j.physletb.2022.137360} {\path{doi:10.1016/j.physletb.2022.137360}}.

\bibitem{Adhikari:2021jrh}
L.~Adhikari, C.~Mondal, S.~Nair, S.~Xu, S.~Jia, X.~Zhao, J.~P. Vary, {Generalized parton distributions and spin structures of light mesons from a light-front Hamiltonian approach}, Phys. Rev. D 104~(11) (2021) 114019.
\newblock \href {http://arxiv.org/abs/2110.05048} {\path{arXiv:2110.05048}}, \href {https://doi.org/10.1103/PhysRevD.104.114019} {\path{doi:10.1103/PhysRevD.104.114019}}.

\bibitem{Zhang:2023xfe}
Z.~Zhang, Z.~Hu, S.~Xu, C.~Mondal, X.~Zhao, J.~P. Vary, {Twist-3 generalized parton distribution for the proton from basis light-front quantization}, Phys. Rev. D 109~(3) (2024) 034031.
\newblock \href {http://arxiv.org/abs/2312.00667} {\path{arXiv:2312.00667}}, \href {https://doi.org/10.1103/PhysRevD.109.034031} {\path{doi:10.1103/PhysRevD.109.034031}}.

\bibitem{Lin:2023ezw}
B.~Lin, S.~Nair, S.~Xu, Z.~Hu, C.~Mondal, X.~Zhao, J.~P. Vary, {Generalized parton distributions of gluon in proton: A light-front quantization approach}, Phys. Lett. B 847 (2023) 138305.
\newblock \href {http://arxiv.org/abs/2308.08275} {\path{arXiv:2308.08275}}, \href {https://doi.org/10.1016/j.physletb.2023.138305} {\path{doi:10.1016/j.physletb.2023.138305}}.

\bibitem{Liu:2022fvl}
Y.~Liu, S.~Xu, C.~Mondal, X.~Zhao, J.~P. Vary, {Angular momentum and generalized parton distributions for the proton with basis light-front quantization}, Phys. Rev. D 105~(9) (2022) 094018.
\newblock \href {http://arxiv.org/abs/2202.00985} {\path{arXiv:2202.00985}}, \href {https://doi.org/10.1103/PhysRevD.105.094018} {\path{doi:10.1103/PhysRevD.105.094018}}.

\bibitem{Jia:2018ary}
S.~Jia, J.~P. Vary, {Basis light front quantization for the charged light mesons with color singlet Nambu\textendash{}Jona-Lasinio interactions}, Phys. Rev. C 99~(3) (2019) 035206.
\newblock \href {http://arxiv.org/abs/1811.08512} {\path{arXiv:1811.08512}}, \href {https://doi.org/10.1103/PhysRevC.99.035206} {\path{doi:10.1103/PhysRevC.99.035206}}.

\bibitem{Chandon2021}
C.~Mondal, S.~Nair, S.~Jia, X.~Zhao, J.~P. Vary, Pion to photon transition form factors with basis light-front quantization, Phys. Rev. D 104 (2021) 094034.
\newblock \href {https://doi.org/10.1103/PhysRevD.104.094034} {\path{doi:10.1103/PhysRevD.104.094034}}.

\bibitem{Brodsky:1997de}
S.~J. Brodsky, H.-C. Pauli, S.~S. Pinsky, {Quantum chromodynamics and other field theories on the light cone}, Phys. Rept. 301 (1998) 299--486.
\newblock \href {http://arxiv.org/abs/hep-ph/9705477} {\path{arXiv:hep-ph/9705477}}, \href {https://doi.org/10.1016/S0370-1573(97)00089-6} {\path{doi:10.1016/S0370-1573(97)00089-6}}.

\bibitem{Glazek:1992aq}
S.~D. Glazek, R.~J. Perry, {Special example of relativistic Hamiltonian field theory}, Phys. Rev. D 45 (1992) 3740--3754.
\newblock \href {https://doi.org/10.1103/PhysRevD.45.3740} {\path{doi:10.1103/PhysRevD.45.3740}}.

\bibitem{Li:2015zda}
Y.~Li, P.~Maris, X.~Zhao, J.~P. Vary, {Heavy Quarkonium in a Holographic Basis}, Phys. Lett. B 758 (2016) 118--124.
\newblock \href {http://arxiv.org/abs/1509.07212} {\path{arXiv:1509.07212}}, \href {https://doi.org/10.1016/j.physletb.2016.04.065} {\path{doi:10.1016/j.physletb.2016.04.065}}.

\bibitem{Brodsky:2014yha}
S.~J. Brodsky, G.~F. de~Teramond, H.~G. Dosch, J.~Erlich, {Light-Front Holographic QCD and Emerging Confinement}, Phys. Rept. 584 (2015) 1--105.
\newblock \href {http://arxiv.org/abs/1407.8131} {\path{arXiv:1407.8131}}, \href {https://doi.org/10.1016/j.physrep.2015.05.001} {\path{doi:10.1016/j.physrep.2015.05.001}}.

\bibitem{Zhao:2014xaa}
X.~Zhao, H.~Honkanen, P.~Maris, J.~P. Vary, S.~J. Brodsky, {Electron g-2 in Light-Front Quantization}, Phys. Lett. B 737 (2014) 65--69.
\newblock \href {http://arxiv.org/abs/1402.4195} {\path{arXiv:1402.4195}}, \href {https://doi.org/10.1016/j.physletb.2014.08.020} {\path{doi:10.1016/j.physletb.2014.08.020}}.

\bibitem{Brodsky2011}
S.~J. Brodsky, F.-G. Cao, G.~F. de~T\'eramond, Evolved qcd predictions for the meson-photon transition form factors, Phys. Rev. D 84 (2011) 033001.
\newblock \href {https://doi.org/10.1103/PhysRevD.84.033001} {\path{doi:10.1103/PhysRevD.84.033001}}.

\bibitem{Ahmady:2022dfv}
M.~Ahmady, S.~Kaur, C.~Mondal, R.~Sandapen, {Pion spectroscopy and dynamics using the holographic light-front Schr\"odinger equation and the 't Hooft equation}, Phys. Lett. B 836 (2023) 137628.
\newblock \href {http://arxiv.org/abs/2208.08405} {\path{arXiv:2208.08405}}, \href {https://doi.org/10.1016/j.physletb.2022.137628} {\path{doi:10.1016/j.physletb.2022.137628}}.

\bibitem{Cao:2021ddi}
F.-G. Cao, {Radial excitation of light mesons and the \ensuremath{\gamma}\ensuremath{\gamma}*\textrightarrow{}\ensuremath{\pi}0 transition form factor}, Phys. Rev. D 104~(5) (2021) 054025.
\newblock \href {http://arxiv.org/abs/2109.02856} {\path{arXiv:2109.02856}}, \href {https://doi.org/10.1103/PhysRevD.104.054025} {\path{doi:10.1103/PhysRevD.104.054025}}.

\bibitem{Mikhailov:2009sa}
S.~V. Mikhailov, N.~G. Stefanis, {Pion transition form factor at the two-loop level vis-a-vis experimental data}, Mod. Phys. Lett. A 24 (2009) 2858--2867.
\newblock \href {http://arxiv.org/abs/0910.3498} {\path{arXiv:0910.3498}}, \href {https://doi.org/10.1142/S0217732309001078} {\path{doi:10.1142/S0217732309001078}}.

\bibitem{Roberts:2010rn}
H.~L.~L. Roberts, C.~D. Roberts, A.~Bashir, L.~X. Gutierrez-Guerrero, P.~C. Tandy, {Abelian anomaly and neutral pion production}, Phys. Rev. C 82 (2010) 065202.
\newblock \href {http://arxiv.org/abs/1009.0067} {\path{arXiv:1009.0067}}, \href {https://doi.org/10.1103/PhysRevC.82.065202} {\path{doi:10.1103/PhysRevC.82.065202}}.

\bibitem{Bakulev:2011rp}
A.~P. Bakulev, S.~V. Mikhailov, A.~V. Pimikov, N.~G. Stefanis, {Pion-photon transition: The New QCD frontier}, Phys. Rev. D 84 (2011) 034014.
\newblock \href {http://arxiv.org/abs/1105.2753} {\path{arXiv:1105.2753}}, \href {https://doi.org/10.1103/PhysRevD.84.034014} {\path{doi:10.1103/PhysRevD.84.034014}}.

\bibitem{Wu:2011gf}
X.-G. Wu, T.~Huang, {Constraints on the Light Pseudoscalar Meson Distribution Amplitudes from Their Meson-Photon Transition Form Factors}, Phys. Rev. D 84 (2011) 074011.
\newblock \href {http://arxiv.org/abs/1106.4365} {\path{arXiv:1106.4365}}, \href {https://doi.org/10.1103/PhysRevD.84.074011} {\path{doi:10.1103/PhysRevD.84.074011}}.

\bibitem{Polyakov:2009je}
M.~V. Polyakov, {On the Pion Distribution Amplitude Shape}, JETP Lett. 90 (2009) 228--231.
\newblock \href {http://arxiv.org/abs/0906.0538} {\path{arXiv:0906.0538}}, \href {https://doi.org/10.1134/S0021364009160024} {\path{doi:10.1134/S0021364009160024}}.

\bibitem{Brodsky:2011yv}
S.~J. Brodsky, F.-G. Cao, G.~F. de~Teramond, {Evolved QCD predictions for the meson-photon transition form factors}, Phys. Rev. D 84 (2011) 033001.
\newblock \href {http://arxiv.org/abs/1104.3364} {\path{arXiv:1104.3364}}, \href {https://doi.org/10.1103/PhysRevD.84.033001} {\path{doi:10.1103/PhysRevD.84.033001}}.

\bibitem{Wu:2010zc}
X.-G. Wu, T.~Huang, {An Implication on the Pion Distribution Amplitude from the Pion-Photon Transition Form Factor with the New BABAR Data}, Phys. Rev. D 82 (2010) 034024.
\newblock \href {http://arxiv.org/abs/1005.3359} {\path{arXiv:1005.3359}}, \href {https://doi.org/10.1103/PhysRevD.82.034024} {\path{doi:10.1103/PhysRevD.82.034024}}.

\bibitem{Kroll:2010bf}
P.~Kroll, {The form factors for the photon to pseudoscalar meson transitions - an update}, Eur. Phys. J. C 71 (2011) 1623.
\newblock \href {http://arxiv.org/abs/1012.3542} {\path{arXiv:1012.3542}}, \href {https://doi.org/10.1140/epjc/s10052-011-1623-4} {\path{doi:10.1140/epjc/s10052-011-1623-4}}.

\bibitem{RuizArriola:2010mrb}
E.~Ruiz~Arriola, W.~Broniowski, {Pion transition form factor in the Regge approach and incomplete vector-meson dominance}, Phys. Rev. D 81 (2010) 094021.
\newblock \href {http://arxiv.org/abs/1004.0837} {\path{arXiv:1004.0837}}, \href {https://doi.org/10.1103/PhysRevD.81.094021} {\path{doi:10.1103/PhysRevD.81.094021}}.

\bibitem{Gorchtein:2011vf}
M.~Gorchtein, P.~Guo, A.~P. Szczepaniak, {Form factors of pseudoscalar mesons}, Phys. Rev. C 86 (2012) 015205.
\newblock \href {http://arxiv.org/abs/1102.5558} {\path{arXiv:1102.5558}}, \href {https://doi.org/10.1103/PhysRevC.86.015205} {\path{doi:10.1103/PhysRevC.86.015205}}.

\bibitem{Pham:2011zi}
T.~N. Pham, X.~Y. Pham, {Chiral Anomaly Effects and the BaBar Measurements of the $\gamma\gamma^{*}\to \pi^{0}$ Transition Form Factor}, Int. J. Mod. Phys. A 26 (2011) 4125--4131.
\newblock \href {http://arxiv.org/abs/1101.3177} {\path{arXiv:1101.3177}}, \href {https://doi.org/10.1142/S0217751X11054140} {\path{doi:10.1142/S0217751X11054140}}.

\bibitem{Dorokhov:2010zzb}
A.~E. Dorokhov, {Photon-pion transition form factor at high photon virtualities within the nonlocal chiral quark model}, JETP Lett. 92 (2010) 707--719.
\newblock \href {https://doi.org/10.1134/S0021364010220145} {\path{doi:10.1134/S0021364010220145}}.

\bibitem{Agaev:2010aq}
S.~S. Agaev, V.~M. Braun, N.~Offen, F.~A. Porkert, {Light Cone Sum Rules for the pi0-gamma*-gamma Form Factor Revisited}, Phys. Rev. D 83 (2011) 054020.
\newblock \href {http://arxiv.org/abs/1012.4671} {\path{arXiv:1012.4671}}, \href {https://doi.org/10.1103/PhysRevD.83.054020} {\path{doi:10.1103/PhysRevD.83.054020}}.

\bibitem{Kotko:2009ij}
P.~Kotko, M.~Praszalowicz, {Covariant Non-local Chiral Quark Model and Pion-photon Transition Distribution Amplitudes}, Phys. Rev. D 80 (2009) 074002.
\newblock \href {http://arxiv.org/abs/0907.4044} {\path{arXiv:0907.4044}}, \href {https://doi.org/10.1103/PhysRevD.80.074002} {\path{doi:10.1103/PhysRevD.80.074002}}.

\end{thebibliography}
\end{document}